\documentclass[journal]{IEEEtran}
\usepackage{mathrsfs}
\usepackage{tabu}
\usepackage{bbm}
\usepackage{amsmath,amscd,amssymb,latexsym}
\usepackage[all]{xy}
\usepackage{pgf,tikz}

\newtheorem{theorem}{Theorem}[section]
\newtheorem{lemma}[theorem]{Lemma}

\newtheorem{example}[theorem]{Example}

\newtheorem{remark}[theorem]{Remark}

\newcommand{\Tr}{\operatorname{Tr}}

\ifCLASSINFOpdf
\else
\fi
\begin{document}
%
\title{ Binary optimal linear codes from posets of the disjoint union of two chains }
%
%
%

\author{Yansheng Wu,~
     Jong Yoon Hyun~ and Qin Yue~

}

\maketitle

\begin{abstract} Recently, Chang and Hyun obtained some classes of  binary optimal codes via  simplicial complexes.
In this letter, we utilize posets of the disjoint union of two chains to construct binary optimal  linear codes. 


\end{abstract}


%
\IEEEpeerreviewmaketitle

\section{Introduction}
%
%
%
%
\IEEEPARstart{L}{et} $\mathbb{F}_2$ be the finite field with two elements and let $\mathbb{F}^n_2$ be a vector space over $\mathbb{F}_2$. 
A  linear code $\mathcal{C}$ of length $n$ over $\Bbb F_2$ is a subspace of $\Bbb F_2^n$
of dimension $k$. If $\mathcal{C}$ has minimal Hamming distance $d $, then $\mathcal{C}$ is called an $[n, k, d]$-code

The weight of $c \in \mathcal{C}$ is the
number of nonzero coordinates of $c$. For $i = 0, 1, \ldots, n$, the numbers $A_i$ denote the frequencies of codewords in the code $\mathcal{C}$ with weight $i$. The weight distribution of a code $\mathcal{C}$ is defined by the sequence $\{A_i\}_{i=0}^n$. Moreover, a code $\mathcal{C}$ is $t$-weight if the sequence $\{A_i\}_{i=1}^n$ has only $t$ nonzero $A_{i}$. If there is no $[n,k,d+1]$ code, then 
 an $ [n,k,d]$ code  $\mathcal{C}$ is called (distance) optimal; if an $[n,k,d+1]$ code   is optimal, then  an $ [n,k,d]$ code  $\mathcal{C}$ is called almost optimal, see \cite[Chapter 2]{HP}.



 For a binary $[n, k, d]$  linear code, the well-known Griesmer bound \cite{G}  is stated as follows: 
 \begin{eqnarray*}n\ge \sum_{i=0}^{k-1}\lceil {\frac{d}{2^i}} \rceil,  \end{eqnarray*} 
 where $\lceil {x} \rceil$ is the ceiling function. We say a linear code is a Griesmer code if the Griesmer bound holds. Note that Griesmer codes are optimal.

 Let $p$ be a prime number, $q$ be a power of  $p$, and $w$ be a power of  $q$. Let $\Bbb F_{w}$ and $\Bbb F_{q}$ be the finite fields with $w$ and $q$ elements, respectively.
 Let $D=\{d_{1}, d_{2}, \ldots, d_{n}\}\subseteq \Bbb F_{w}$ and $\Tr_{w/q}$ be the trace function from $\Bbb F_{w}$ to $\Bbb F_{q}$. Ding et al. \cite{D1, DN} first defined a linear code  over $\Bbb F_{q}$ of length $n$
\begin{equation}
\mathcal{C}_{D}= \{(\Tr_{w/q}(xd_{1}), \ldots, \Tr_{w/q}(xd_{n})) : x\in  \Bbb F_{w}\}.
\end{equation}
 Many known codes with good parameters could be obtained by some suitable defining sets $D$, see \cite{HY1}-\cite{LL}. 
 


In this letter, we turn our interest to a typical construction \cite{HP} of a linear code. 
Let $D=\{ g_1, g_2,\ldots, g_n \}\subseteq \Bbb F_p^m$. A $p$-ary linear code with length $n$  is defined by
\begin{equation}
\mathcal{C}_{D}= \{c_{u}=(u\cdot g_1, u\cdot  g_2, \ldots, u\cdot { g}_n): { u}\in  \Bbb F_{p}^m\}, 
\end{equation}
where $u\cdot v$ denotes  Euclidean inner product of $u$ and $v$ in $\Bbb F_p^m.$ 

Based on generic constructions of linear codes, Chang and Hyun \cite{CH} constructed  binary optimal and minimal   linear codes  via some  simplicial complexes. The main idea of this letter is obtaining some binary optimal codes  by employing  posets of the disjoint union of two chains. Especially, anti-chains are corresponded to simplicial complexes. Section II introduces the concept of posets. Specially, we determine the down sets of the disjoint union of two chains.
Section III presents the weight distributions of the binary linear codes under the case of the disjoint union of two chains. Section IV obtains some binary optimal linear codes. Section V concludes this paper.



\section{Disjoint union of two chains}

Throughout this letter, let $\mathbb{P}=([n], \preceq)$ be a partially ordered set (simply, poset) on $[n]=\{1,2,\ldots,n\}$ equipped with order relation $\preceq$. A subset $I$ of $\mathbb{P}$ is called a down set of $\mathbb{P}$ if $x$ is in $I$ and $y \preceq x$ imply $y$ is in $I$. We use $\mathcal{L}(\mathbb{P})$ to denote the set of down sets of $\mathbb{P}$. For a down set $I$ of $\mathbb{P}$, by $I(\mathbb{P})$ we mean the set of down sets of $\mathbb{P}$ which is contained in $I$. We can readily check that $I(\mathbb{P})$ is a down set of $\mathcal{L}(\mathbb{P})$. 

Two distinct elements $i$ and $j$ in $[n]$ are called comparable if either $i\preceq j$ or $j\preceq i$, and incomparable otherwise. It is said that a poset $\mathbb{P}$ is a chain (resp., an anti-chain) if every pair of distinct elements is comparable (resp. incomparable), see  \cite{N}. 

A poset $\mathbb{P}$ can be represented by a Hasse diagram, which represents each elements of the poset $\mathbb{P}$ by a distinct point; the point $y$ is situated {\it higher than} the point  $x$ if $x\prec y$.

For $m\in [n]$, we denote by $\mathbb{P}=(m\oplus n, \preceq)$ the disjoint union of two chains that is given by the Hasse diagram in Figure 1.

\begin{center}
\begin{tikzpicture}

\draw (0, -1)[fill = black] circle (0.05);
\draw (0, -2)[fill = black] circle (0.05);
\draw (0, -2.5)[fill = black] circle (0.05);
\draw (0, -3)[fill = black] circle (0.05);
\draw (0, -3.5)[fill = black] circle (0.05);
\draw (0, -4)[fill = black] circle (0.05);
\draw (0, -5)[fill = black] circle (0.05);

\draw (3, -1)[fill = black] circle (0.05);
\draw (3, -2)[fill = black] circle (0.05);
\draw (3, -2.5)[fill = black] circle (0.05);
\draw (3, -3)[fill = black] circle (0.05);
\draw (3, -3.5)[fill = black] circle (0.05);
\draw (3, -4)[fill = black] circle (0.05);
\draw (3, -5)[fill = black] circle (0.05);

\draw[thin] (0, -5) --(0,-4); 

\draw[thin] (0,-2)--(0,-1);
\draw[thin] (3, -5) --(3,-4); 

\draw[thin] (3,-2)--(3,-1);

\node  at (0, -1) [left] {$~~~m~~$};
\node  at (0, -2) [left] {$~~~m-1~~$};
\node  at (0, -4) [left] {$~~~2~~$};
\node  at (0, -5) [left] {$~~~1~~$};
\node  at (3, -4) [right] {$~~m+2~~$};
\node  at (3, -5) [right] {$~~m+1~~$};
\node  at (3, -1) [right] {$~~~n~~$};\node  at (3, -2) [right] {$~~~n-1~~$};

\node  at (2, -5.5) [below] {Figure 1 $\mathbb{P}=$($m\oplus n, \preceq$)};

\end{tikzpicture}
\end{center}

For a vector $v \in \Bbb F_2^n$, the support $\mathrm{supp}(v)$  is defined by the set $\{1\le i\le n: v_i \neq0\}$. Then the  weight $wt(v)$ of $v\in \mathbb{F}^n_2$ satisfies that
$wt(v) = |\mathrm{supp}(v)|$. We express the power set of $[n]$ by $2^{[n]}$. Then there is a bijection between $\mathbb{F}_2^n$ and $2^{[n]}$, defined by $v\mapsto$ supp$(v)$. Throughout this letter, we should identify a vector in $\mathbb{F}_2^n$  with its support.


Let $X$ be a collection of $2^{[n]}$.
Define 
$$\mathcal{H}_{X}(x_1,x_2\ldots, x_n)=\sum_{u\in X}\prod_{i=1}^nx_i^{u_i}\in \mathbb{Z}[x_1,x_2, \ldots, x_n],
$$
where $u=(u_1,u_2,\ldots, u_n)\in \mathbb{F}_2^n$ and $\mathbb{Z}$ is the ring of integers. Notice that $\mathcal{H}_{\emptyset}(x_1,x_2\ldots, x_n)=0$ by convention,
and $|X|=\mathcal{H}_{X}(1,1\ldots, 1)$.


By Figure 1, the following two lemmas are easy to verify.  
\begin{lemma} Let $i$ and $j$ are two integers with $1\le i\le m$ and $m+1\le j\le n$. A down set of $\mathbb{P}=(m\oplus n, \preceq)$ must be one of the following forms:

$(i)$ $[i]$.

$(ii)$ $[j]\backslash [m]$.

$(iii)$ $[i] \cup ([j]\backslash [m])$.
\end{lemma}

\begin{lemma} Suppose that $I$ is a down set of $\mathbb{P}=(m\oplus n, \preceq)$. 

$(i)$ If $I=[i]$, then $$\mathcal{H}_{I(\mathbb{P})}(x_1,x_2\ldots, x_n)=1+\sum_{k=1}^ix_1\cdots x_k.$$ 

$(ii)$ If $I=[j]\backslash [m]$, then $$\mathcal{H}_{I(\mathbb{P})}(x_1,x_2\ldots, x_n)=1+\sum_{l=m+1}^jx_{m+1}\cdots x_l.$$ 

$(iii)$  If $I=[i] \cup ([j]\backslash [m])$, then \begin{eqnarray*}&&\mathcal{H}_{I(\mathbb{P})}(x_1,x_2\ldots, x_n)\\
&=&1+\sum_{k=1}^ix_1\cdots x_k+\sum_{l=m+1}^jx_{m+1}\cdots x_l\\
&+&\sum_{k=1}^i \sum_{l=m+1}^j x_1\cdots x_kx_{m+1}\cdots x_l .\end{eqnarray*} 
\end{lemma}

\section{Weight distributions of binary linear codes}

Assume that  $I$ is a down set of $\mathbb{P}=(m\oplus n, \preceq)$ in Lemma 2.1. In Eq. (2), let $D=\Bbb F_2^n \backslash I(\mathbb{P})$.  

\begin{theorem} 
(I) If $I=[i]$ and $1\le i\le m$, then the weight distribution of  $\mathcal{C}_D$ is presented in Table I with parameters $[2^n-i-1,n]$.
\begin{table} 
\caption{Theorem 3.1 (i)}  
\begin{center}  
\begin{tabu} to 0.35\textwidth{X[2,c]|X[1,c]} 
\hline 
Weight &Frequency\\ 
\hline 
$0$&$1$\\ 
 \hline 
$2^{n-1}-i+s~(0\le s< i)$&$2^{n-i}{i\choose s}$\\ 
\hline 
$2^{n-1}$&$2^{n-i}-1$
\\
\hline 
\end{tabu}  
\end{center}  
\end{table}

(II) If $I=[j]\backslash [m]$ and $m+1\le j\le n$, then the weight distribution of  $\mathcal{C}_D$ is presented in Table II with
 parameters $[2^n+m-j-1,n]$.
\begin{table} [h]
\caption{Theorem 3.1 (ii)}  
\begin{center}  
\begin{tabu} to 0.4\textwidth{X[1.9,c]|X[1,c]} 
\hline 
Weight &Frequency\\ 
\hline 
$0$&$1$\\ 
 \hline 
$2^{n-1}-(j-m)+t~ (0\le t< j-m)$&$2^{n-(j-m)}{j-m\choose t}$\\ 
\hline 
$2^{n-1}$&$2^{n-(j-m)}-1$
\\
\hline 
\end{tabu}  
\end{center}  
\end{table}

(III)  If $I=[i] \cup ([j]\backslash [m])$, $1\le i\le m$, and  $m+1\le j\le n$, then the weight distribution of $\mathcal{C}_D$ is presented in Table III with parameters $[2^n+m-i-j-1-i(j-m),n]$.
\begin{table} [h]
\caption{Theorem 3.1 (iii)}  
\begin{center}  
\begin{tabu} to 0.5\textwidth{X[2,c]|X[1,c]} 
\hline 
Weight &Frequency\\ 
\hline 
$0$&$1$\\ 
 \hline 
 $2^{n-1}+s+t+2st-(s+1)(j-m)-(t+1)i $&\\
 $0\le s\le i, 0\le t\le (j-m),$&$2^{n-i-(j-m)}{i\choose s}{j-m\choose t}$\\
 $(s,t)\neq (i, j-m)$&\\
\hline 
$2^{n-1}$&$2^{n-i-(j-m)}-1$
\\
\hline 
\end{tabu}  
\end{center}  
\end{table}
\end{theorem}

{\bf Proof} 
The length of $\mathcal{C}_D$ is $2^n-|I(\mathbb{P})|$. For $u=(u_1,u_2,\ldots,u_n)\in\mathbb{F}^n_2$,  the Hamming weight of $c_{ u}\in \mathcal{C}_D$ is   
\begin{eqnarray}
&&|D|-\frac12\sum_{y\in\Bbb F_2}\sum_{x\in D}(-1)^{(u\cdot x)y}\nonumber\\
&=&\frac {|D|}2-\frac12\sum_{x\in D}(-1)^{(u\cdot x)}\nonumber\\
&=&\frac {|D|}2+\frac12\sum_{x\in {I}(\mathbb{P})}(-1)^{u_1 x_1}(-1)^{u_2 x_2}\cdots(-1)^{u_n x_n}\nonumber\\
&=&\frac {|D|}2+\frac12\mathcal{H}_{{I}(\mathbb{P})}((-1)^{u_1},(-1)^{u_2},\ldots, (-1)^{u_n}).
\end{eqnarray}
For any $u=(u_1,u_2,\ldots, u_n)\in \mathbb{F}_2^n$, write $u=(v,w)$, where $v=(u_1,\ldots, u_m)$ and $w=(u_{m+1},\ldots, u_n)$. Let $\overline{v}_k=(u_1, \ldots, u_k)$, where $1\le k\le i$ and $k$ is an integer. Assume that the number of such $k$ satisfying $wt(\overline{v}_k)\equiv 0 \pmod 2$ is $s$.

(I) If $I=[i]$, then  $|D|=2^n-(i+1)$. By Lemma 2.2 and Eq. (3),   
\begin{eqnarray*}
&&wt(c_{ u})=2^{n-1}-\frac{i+1}{2}+\frac12(1+\sum_{k=1}^i(-1)^{u_1+\cdots+u_k})\\
&=&2^{n-1}-\frac{i}{2}+\frac12\sum_{k=1}^i(-1)^{wt(\overline{v}_k)}\\
&=&2^{n-1}-\frac{i}{2}+\frac12(s-(i-s))= 2^{n-1}-i+s. 
\end{eqnarray*}

 If $s=i$, i.e., $\overline{v}_i=(0, \ldots,0)$, then there are $2^{n-i}-1$ nonzero vectors $u\in \Bbb F_2^n$ such that  $wt(c_u)=2^{n-1}$. 
 
 If $0\le s\le i$, then there are  $2^{n-i}{i\choose s}$ nonzero vectors $ u\in \Bbb F_2^n$ such that $wt(c_u)=2^{n-1}-i+s$. Hence we have the result in Table I.

(II) If $I=[j]\backslash [m]$, then $|D|=2^n-(j-m+1)$.
For any ${ 0}\neq u=(u_1,u_2,\ldots, u_n)\in \mathbb{Z}_2^n$, write $u=(v,w)$, $v=(u_1,\ldots, u_m)$,  and $w=(u_{m+1},\ldots, u_n)$. Let $\overline{w}_l=(u_{m+1}, \ldots, u_l)$, where $l$ is an integer with $m+1\le l\le j$. Assume that the number of such $l$ satisfying $wt(\overline{w}_l)\equiv 0 \pmod 2$ is $t$. By Lemma 2.2 and Eq. (3),   
\begin{eqnarray*}&&wt(c_{ u})\\
&=&2^{n-1}-\frac{j-m+1}{2}+\frac12(1+\sum_{l=m+1}^j(-1)^{u_{m+1}+\cdots+u_l})\\
&=&2^{n-1}-\frac{j-m}{2}+\frac12\sum_{l=m+1}^j(-1)^{wt(\overline{w}_l)}\\
&=&2^{n-1}-\frac{j-m}{2}+\frac12(t-(j-m-t))\\
&=& 2^{n-1}-j+m+t. \end{eqnarray*}

 If $t=j-m$, i.e., $\overline{w}_{j-m}=(0,\ldots, 0)$, then there are $2^{n+m-j}-1$ nonzero vectors $u\in \Bbb F_2^n$ such that $wt(c_u)=2^{n-1}$. 
 
 If $0\le t\le j-m$,  then there are  $2^{n+m-j}C_{j-m}^t$  nonzero vectors $u\in \Bbb F_2^n$ such that $wt(c_u)=2^{n-1}-j+m+t$. Hence we have the result in Table II. 

(III) If $I=[i] \cup ([j]\backslash [m])$, then $|D|=2^n-(i+1)-(j-m+1)+1-i(j-m)=2^n+m-i-j-1-i(j-m)$. By the proofs of $(i)$, Lemma 2 and Eq. (3), 
\begin{eqnarray*}
&&wt(c_{ u})=2^{n-1}-\frac{i+j+1-m+i(j-m)}{2}\\
&+&\frac12(1+\sum_{k=1}^i(-1)^{u_1+\cdots+u_k}+\sum_{l=m+1}^j(-1)^{u_{m+1}+\cdots+u_l})\\
&+&\frac{1}{2}  (\sum_{k=1}^i(-1)^{u_1+\cdots+u_k}\sum_{l=m+1}^j(-1)^{u_{m+1}+\cdots+u_l})\\
&=&2^{n-1}-\frac{i+j-m+i(j-m)}{2}\\
&+&\frac12(\sum_{k=1}^i(-1)^{wt(\overline{v}_k)}+\sum_{l=m+1}^j(-1)^{wt(\overline{w}_l)})\\
&+ & \frac12(\sum_{k=1}^i(-1)^{wt(\overline{v}_k)}  )(  \sum_{l=m+1}^j(-1)^{wt(\overline{w}_l)})\\
&=&2^{n-1}-\frac{i+j-m+i(j-m)}{2} \\
&+&\frac12[(s-(i-s)) + t-(j-m-t)]\\
&+&\frac12(s-(i-s))(t-(j-m-t))\\
&=&2^{n-1}+s+t+2st-(s+1)(j-m)-(t+1)i.
\end{eqnarray*}

If $s=i$ and $t=j-m$, i.e., $\overline{v}_i=(0,\ldots, 0)$ and $\overline{w}_{j-m}=(0,\ldots, 0)$, then there are $2^{n-i-(j-m)}-1$ nonzero vectors $u\in \Bbb F_2^n$ such that  $wt(c_u)=2^{n-1}$.
 
If $(s,t)\neq (i,j-m)$,  then there are  $2^{n-i-(j-m)}{i\choose s}{t\choose j-m}$ nonzero  vectors $u\in \Bbb F_2^n$ such that $wt(c_u)=2^{n-1}+s+t+2st-(s+1)(j-m)-(t+1)i$. Hence we have the result in Table III. 
$\blacksquare$

\begin{remark} Due to the special structure of poset of the disjoint union of two chains, we just need to consider the case of  a single order ideal. 
\end{remark}

\section{Optimal binary linear codes}

By Theorem 3.1,  we will obtain some  optimal or almost optimal  binary linear codes.

\begin{theorem} 
Let $\mathcal{C}_D$ be the code in Theorem 3.1 (I).

$(i)$ If $n\ge 2$ is an integer and $i=1$, then the code $\mathcal{C}_D$ is a Griesmer code.

$(ii)$ If $n> m\ge 2$ is an integer and $i=2$, then the code $\mathcal{C}_D$ is  optimal.

$(iii)$ If $n\ge 3$ is an integer and $i=3$, then  the code $\mathcal{C}_D$ is almost optimal.
\end{theorem}

{\bf Proof} 
$(i)$ If $i=1$, then by Table I, the minimum distance of the code $\mathcal{C}_{D}$ is $2^{n-1}-1$. Then 
\begin{eqnarray*}&&\sum_{i=0}^{n-1}\bigg\lceil {\frac{2^{n-1}-1}{2^i}} \bigg\rceil =(2^n-1)-(2-1)+1= 2^n-2.
\end{eqnarray*} 
Therefore, the code $\mathcal{C}_{D}$ is a Griesmer code. 

$(ii)$ If $i=2$, then by Table I, the minimum distance of the code  $\mathcal{C}_{D}$ is $2^{n-1}-2$. By the proof of Theorem 4.1, $$\sum_{i=0}^{n-1}\bigg\lceil {\frac{2^{n-1}-1}{2^i}} \bigg\rceil = 2^n-2>2^n-3.$$ Therefore, there is no  $[2^n-3,n,2^{n-1}-1]$ linear code exist. Namely, the code $\mathcal{C}_{D}$ is optimal.

$(iii)$ If $i=3$, then by Table I, $d=2^{n-1}-3$. Then $$\sum_{i=0}^{n-1}\bigg\lceil {\frac{2^{n-1}-3}{2^i}} \bigg\rceil = 2^n-5<2^n-4$$ and $\sum_{i=0}^{n-1}\bigg\lceil {\frac{2^{n-1}-2}{2^i}} \bigg\rceil = 2^n-4$. Therefore, the code with parameters $[2^n-4, n,2^{n-1}-2]$ is optimal. Namely, $\mathcal{C}_D$ is almost optimal.
$\blacksquare$


\begin{theorem}
Let $\mathcal{C}_D$ be the code in Theorem 3.1 (III).

$(i)$ If $n\ge 3$ is an integer, $m=2$, $i=1$ and $j=3$, then the code $\mathcal{C}_D$ is a  Griesmer code. 

$(ii)$ If $n\ge 4$ is an integer, $m=2$,  $i=2$ and $j=3$, then the code $\mathcal{C}_D$ is optimal. 

$(iii)$ If $n\ge 4$ is an integer, $m\ge i=2$ and $j-m=2$, then the code $\mathcal{C}_D$ is  almost optimal. 
\end{theorem}
 {\bf Proof} 
 
$(i)$ The result follows from  $\sum_{i=0}^{n-1}\bigg\lceil {\frac{2^{n-1}-2}{2^i}} \bigg\rceil = 2^n-4$.

$(ii)$ The result follows from  $\sum_{i=0}^{n-1}\bigg\lceil {\frac{2^{n-1}-3}{2^i}} \bigg\rceil = 2^n-5>2^n-6$.

$(iii)$ Note that $\sum_{i=0}^{n-1}\bigg\lceil {\frac{2^{n-1}-5}{2^i}} \bigg\rceil = 2^n-9$. That means a $[2^n-9, n, 2^{n-1}-5]$ is Griesmer code. Hence the code $\mathcal{C}_D$ is almost optimal.
  $\blacksquare$

  \begin{example} Let $n=6$, $m=4$ and $I=[4]$. Then $D=\Bbb F_2^6 \backslash I(\mathbb{P})=\Bbb F_2^6\backslash \{ 0, v_1, v_2, v_3, v_4\}$, where $$v_4=(1,1,1,1,0,0) ,v_3=(1,1,1,0,0,0)  , $$   $$v_2= (1,1,0, 0,0,0), v_1=(1, 0, 0, 0,0,0).$$ By Theorem 3.1, the code $\mathcal{C}_D$ has parameters $[59,6, 28]$ and its weight enumerator is given by $$1+4z^{28}+16z^{29}+24z^{30}+16z^{31}+3z^{32}.$$ 
  This result is confirmed by Magma. According to  \cite{G2},  the optimal linear code has parameters $[59,6,29]$ .
  \end{example}

    \begin{example} Let $n=6$, $m=4$, $i=3$, $j=6$, and $I=[3]\cup ([6]\backslash  [4])=\{1,2,3, 5,6\}$. Then $D=\Bbb F_2^6 \backslash I(\mathbb{P})=\Bbb F_2^6\backslash \{ 0, v_1, v_2, v_3, v_4, v_5,v_6, v_7, v_8, v_9, v_{10}, v_{11}\}$, where $$v_1=(1, 0, 0, 0,0,0),v_2= (1,1,0,0, 0,0),$$ $$v_3=(1,1,1,0,0,0),v_4=(0,0,0,0,1,0), $$  $$v_5=(0,0,0,0,1,1),v_6=(1,0,0,0,1,0), $$ $$v_7=(1,0,0,0,1,1),v_8=(1,1,0,0,1,0), $$ $$v_9=(1,1,0,0,1,1),v_{10}=(1,1,1,0,1,0), $$ $$v_{11}=(1,1,1,0,1,1).$$
    By Theorem 3.1, the code $\mathcal{C}_D$ has parameters $[52,6, 23]$  and its weight enumerator  is given by $$1+2z^{23}+2z^{24}+10z^{25}+24z^{26}+14z^{27}+4z^{28}+6z^{29}+z^{32}.$$
    This result is confirmed by Magma. According to \cite{G2},  the optimal linear code has parameters $[52,6,25]$ . 
  \end{example}

\section{Conclusion}

By using the disjoint union of two chains, we obtained several families of binary optimal and almost optimal linear codes based on the generic construction. More binary optimal linear codes could be found from the code  in Theorem 3.1.





%




\ifCLASSOPTIONcaptionsoff
  \newpage
\fi

\end{document}